\documentclass{PoS}
\usepackage{amsmath}
\usepackage[dvipdfmx]{}

\title{
\begin{picture}(0,0)(0,0)%
   \put(125,60){\makebox(0,0)[l]{\textnormal{\normalsize 
   RIKEN-QHP-206, YITP-15-100, KEK-CP-335, KYUSHU-HET-159 }}}%
   \end{picture}%
Thermodynamics and reference scale of SU(3) gauge theory from gradient flow on fine lattices}

\ShortTitle{Thermodynamics and reference scale from gradient flow}

\author{\speaker{Masakiyo Kitazawa}\\
        Department of Physics, Osaka University, 
        Toyonaka, Osaka 560-0043, Japan\\
        E-mail: \email{kitazawa@phys.sci.osaka-u.ac.jp}}

\author{Masayuki Asakawa\\
        Department of Physics, Osaka University, 
        Toyonaka, Osaka 560-0043, Japan\\
        E-mail: \email{yuki@phys.sci.osaka-u.ac.jp}}

\author{Tetsuo Hatsuda\\
        Theoretical Research Division, Nishina Center, RIKEN, Wako
        351-0198, Japan,\\
        Kavli IPMU (WPI), The University of Tokyo, Chiba 606-8502, Japan \\
        E-mail: \email{thatsuda@riken.jp}}

\author{Takumi Iritani\\
        Yukawa Institute for Theoretical Physics, Kyoto
        606-8512, Japan;\\
        Department of Physics and Astronomy, Stony Brook University, 
        Stony Brook, New York 11794-3800, USA \\
        E-mail: \email{iritani@yukawa.kyoto-u.ac.jp}, \email{takumi.iritani@stonybrook.edu}}

\author{Etsuko Itou\\
        High Energy Accelerator Research Organisation (KEK), Tsukuba
        305-0801, Japan\\
        E-mail: \email{eitou@post.kek.jp}}

\author{Hiroshi Suzuki\\
        Department of Physics, Kyushu University, 744 Motooka, 
        Nishi-ku, Fukuoka, 812-0395, Japan\\
        E-mail: \email{hsuzuki@phys.kyushu-u.ac.jp}}

\abstract{
We study the parametrization of lattice spacing and thermodynamics of 
SU(3) gauge theory on the basis of the Yang-Mills gradient flow 
on fine lattices. 
The lattice spacing of the Wilson gauge action 
is determined over a wide range $6.3\le\beta\le7.5$ with high accuracy. 
The measurements of the flow time and lattice spacing dependences of 
the expectation values of the energy-momentum tensor are performed
on fine lattices.
}

\FullConference{The 33rd International Symposium on Lattice Field Theory\\
		14 -18 July 2015\\
		Kobe International Conference Center, Kobe, Japan*}

\begin{document}

\section{Introduction}

After the introduction of the Yang-Mills gradient flow 
in lattice gauge theory \cite{Luscher:2010iy}, this concept 
has been successfully applied to various purposes in lattice QCD 
numerical simulations \cite{review1,review2}.
In this proceedings, we report our recent studies on the 
application of the gradient flow for two purposes.

In the first analysis, we measure the lattice spacing of the SU(3) 
Wilson gauge action \cite{Asakawa:2015vta}.
We introduce reference scales having physical dimension on the 
basis of the gradient flow \cite{Luscher:2010iy,Borsanyi:2012zs},
and perform the measurement of lattice spacing in this scale
in the range $6.3\le\beta=6/g^2\le7.5$.
We use the reference scale called $w_{0.4}$ in this analysis.
The parametrization of the lattice spacing as a function of 
$\beta$ is given in Eq.~(\ref{eq:fitfinal}).
The relation between $w_{0.4}$ and other reference scales, 
as well as the finite volume and lattice discretization effects 
in our analysis are also discussed \cite{Asakawa:2015vta}.

Secondly, we apply the gradient flow for the measurement of 
thermodynamics of SU(3) gauge theory.
In this analysis, we use the energy-momentum tensor (EMT) operator
defined by the gradient flow \cite{Suzuki:2013gza} 
with the small flow-time expansion \cite{Luscher:2011bx}.
The thermodynamic quantities are obtained by taking the expectation 
values of the EMT operator.
In Ref.~\cite{Asakawa:2013laa}, we performed the measurement 
of the thermodynamics of SU(3) gauge theory, and found that 
this method can successfully analyze the thermodynamics with 
good statistics compared with the previously-known integral 
method.
In this proceedings we report the update of this analysis
on finer lattices with temporal lattice size $N_t=12$--$32$.

\section{Gradient flow}

Let us first give a brief review on the gradient flow.
The gradient flow for the Yang-Mills gauge field is the 
continuous transformation of the field defined by the 
differential equation \cite{Luscher:2010iy}
\begin{align}
\frac{d A_\mu}{dt} = - g_0^2 
\frac{ \partial S_{\rm YM}(t)}{ \partial A_\mu }
= D_\nu G_{\nu\mu} ,
\label{eq:GF}
\end{align}
with the Yang-Mills action $S_{\rm YM}(t)$ composed of $A_\mu(t)$.
Color indices are suppressed for simplicity.
The initial condition at $t=0$ is taken for the field in the
conventional gauge theory; $A_\mu(0)=A_\mu$.
The flow time $t$, which has a dimension of inverse mass squared, 
is a parameter which controls the transformation.
The gauge field is transformed along the steepest descent 
direction as $t$ increases.
At the tree level, Eq.~(\ref{eq:GF}) is rewritten as 
\begin{align}
  \frac{d A_\mu}{dt} = \partial_\nu \partial_\nu A_\mu
  + {\rm (gauge ~ dependent ~ terms)}.
\label{eq:diffusion}
\end{align}
Neglecting the gauge dependent terms, Eq.~(\ref{eq:diffusion})
is the diffusion equation in four-dimensional space.
For positive $t$, therefore, the gradient flow acts as 
the cooling of the gauge field with the smearing radius 
$\sqrt{8t}$.

In Ref.~\cite{Luscher:2011bx}, it is rigorously proved that all 
composite operators composed of $A_\mu(t)$ take finite values for $t>0$.
This property ensures that observables at $t>0$ are 
regularization independent.
In the present study, we use the gradient flow for two purposes;
(1) the introduction of reference scales and the measurement of
lattice spacing \cite{Luscher:2010iy}, and 
(2) the measurement of thermodynamic observables \cite{Asakawa:2013laa}
using the energy-momentum tensor operator defined by the 
gradient flow with the small flow-time expansion \cite{Suzuki:2013gza}.

In the present study we consider two observables,
\begin{align}
  E(t) &= \frac14 G_{\mu\nu}^a(t)G_{\mu\nu}^a(t),
  \label{eq:E}
\\
  U_{\mu\nu}(t,x) &= G_{\mu\rho}^a (t,x)G_{\nu\rho}^a (t,x)
  -\frac14 \delta_{\mu\nu}G_{\rho\sigma}^a(t,x)G_{\rho\sigma}^a(t,x), 
  \label{eq:U}
\end{align}
at positive flow time $t>0$,
where $G_{\mu\nu}^a(t)$ is the ``field strength'' composed of~$A_\mu(t)$.

\section{Reference scale and lattice spacing}

In this section, we analyze the lattice spacing of SU(3) Wilson
gauge action \cite{Asakawa:2015vta}.
In order to define reference scales and measure lattice
spacing with the gradient flow, we use an observation that 
the expectation value of an observable at $t>0$ is regularization
independent \cite{Luscher:2010iy}.
From this property, the expectation value of dimensionless
observables does not depend on regularization.
On the lattice, therefore, they should be the lattice spacing, $a$,
independent up to $O(a^2)$ effects.
By choosing $t^2 E(t)$ as the dimensionless observable,
the value of~$t$ at which $t^2\langle E(t)\rangle$ takes a specific 
value~$X$, i.e. the solution of the equation
\begin{equation}
   \left.t^2\langle E(t)\rangle\right|_{t=t_{_X}}=X,
\label{eq:t_x}
\end{equation}
is a dimensionful quantity, which can be used as a reference scale 
to introduce physical unit in lattice gauge theory. 
In~Ref.~\cite{Luscher:2010iy},
$t_{_{X}=0.3}$ (sometimes called~$t_0$) is used as the reference scale.
In~Ref.~\cite{Borsanyi:2012zs}, a quantity~$w_{_X}$ defined by 
\begin{equation}
   \left.t\frac{d}{dt}t^2\langle E(t)\rangle\right|_{t=w_{_X}^2}=X,
\label{eq:w_x}
\end{equation}
is proposed as an alternative reference scale. In~Ref.~\cite{Borsanyi:2012zs},
a reference scale~$w_{_{X}=0.3}$ (sometimes called~$w_0$) is employed to set the
scale.

In the present study we consider reference scales, $t_{_X}$
and~$w_{_X}$ with $X=0.2$, $0.3$ and~$0.4$.
Larger $X$ is preferable to suppress the lattice discretization
error, while the smearing radius~$\sqrt{8t}$ would
eventually hit the lattice boundary for too large~$X$. 
We use $w_{0.4}$ and~$w_{0.2}$ for the reference scales and introduce a
new parametrization of the lattice spacing~$a$ in terms of the bare
coupling~$\beta=6/g_0^2$ by a hybrid use of these reference scales.
In this method, we measure the lattice spacing in the range 
$6.3<\beta<7.5$.

\begin{table}[t]
\begin{center}
\begin{tabular}{|c||c|c|c|c|c|c|c|c|c|c|c|}
\hline $\beta$ & 
6.3 & 6.4 & 6.5 & 6.6 & 6.7 & 6.8 & 6.9 & 7.0 & 7.2 & 7.4 & 7.5 \\
\hline $N_{\rm s}$ & 
 64 &  64 &  64 &  64 &  64 &  64 &  64 &  96 &  96 & 128 & 128 \\
\hline $N_{\rm conf}$ &
 30 & 100 &  49 & 100 &  30 & 100 &  30 &  60 &  53 &  40 &  60 \\
\hline
\end{tabular}
\caption{
Simulation parameters $\beta=6/g_0^2$, the lattice size
$N_{\rm s}$ and the number of configurations $N_{\mathrm{conf}}$
\cite{Asakawa:2015vta}.
}
\label{table:param}
\end{center}
\end{table}

We perform numerical analyses of the SU(3) Yang--Mills
theory with the Wilson plaquette action with 
the periodic boundary condition with the lattice size~$N_{\rm s}^4$. 
The values of~$\beta=6/g_0^2$, $N_{\rm s}$ and the number of 
configurations $N_{\mathrm{conf}}$ are summarized
in~Table~\ref{table:param}.
For $\beta=7.0$, $7.2$ and $7.4$, we have performed another measurements
with different spatial sizes; $N_{\rm s}=64$, $64$ and $96$, for 
$\beta=7.0$, $7.2$ and $7.4$, respectively.
From the comparison of the numerical results with different $N_{\rm s}$,
we have checked that the finite volume effects are well suppressed
with the choices of $N_{\rm s}$ shown in Table~\ref{table:param}
\cite{Asakawa:2015vta}.

We use the Wilson
gauge action~$S_{\mathrm{YM}}$ for the flow equation
in~Eq.~(\ref{eq:GF}). To construct the operator~$E$, we
use the clover-type representation of~$G_{\mu\nu}^a$ 
unless otherwise stated.
The numerical error of the Runge--Kutta (RK) method to solve 
the differential equation Eq.~(\ref{eq:GF}) have been estimated by 
comparing the numerical results with different integration step sizes.
We have checked that the numerical error of the RK method is 
within two orders in magnitude smaller than the statistical errors.

Autocorrelation between different configurations is analyzed 
by the autocorrelation function and the dependence of the jackknife 
statistical errors against the bin-size, $N_{\mathrm{bin}}$.
These analyses show that the autocorrelation is not visible within 
statistics in our set of gauge configurations, which are separated 
by $1,000$ Monte-Carlo steps composed of heatbath and five 
over-relaxation updates.
On the other hand, 
the autocorrelation of the topological charge
is known to become longer as the lattice spacing becomes
finer due to the critical slowing down \cite{Luscher:2010iy}.
It is desirable to perform new measurements of the topological 
charge to investigate the effect of the critical slowing down
on our analysis, which is left for our future work. 

As pointed out in~Ref.~\cite{Borsanyi:2012zs}, the discretization error
of~$w_{0.3}$ is smaller than $t_{0.3}$.
We thus employ $w_{_X}$ as the key reference scale in this paper. 
We also note that the lattice artifact is expected to 
be smaller for larger~$X$.
To check the discretization effects, 
we compare $t^2\langle E(t)\rangle$ and 
$t\frac{d}{dt}t^2\langle E(t)\rangle$ defined from the clover-type 
representation with those defined from $E(t)=2(1-P(t))$ using 
the average plaquette~$P(t)$ \cite{Asakawa:2015vta}.
This analysis shows that the difference between the two definitions 
is suppressed for large~$t$. Moreover, the difference is more 
suppressed in~$t\frac{d}{dt}t^2\langle E(t)\rangle$ than
that in~$t^2\langle E(t)\rangle$, i.e. the discretization effect 
in the former is smaller than the latter.
We thus employ $w_{0.4}$ as the key reference scale. 
In our simulations, we estimate $w_{0.4}/a$ for $\beta=7.4$, $7.5$
using the data at small flow time ($w_{0.2}/a$) for these $\beta$ 
and the extrapolation of the ratio $w_{0.4}/w_{0.2}=1.3042(9)$ 
to the continuum limit \cite{Asakawa:2015vta}.

\begin{figure}[t]
\begin{center}
\includegraphics[width=0.6\textwidth]{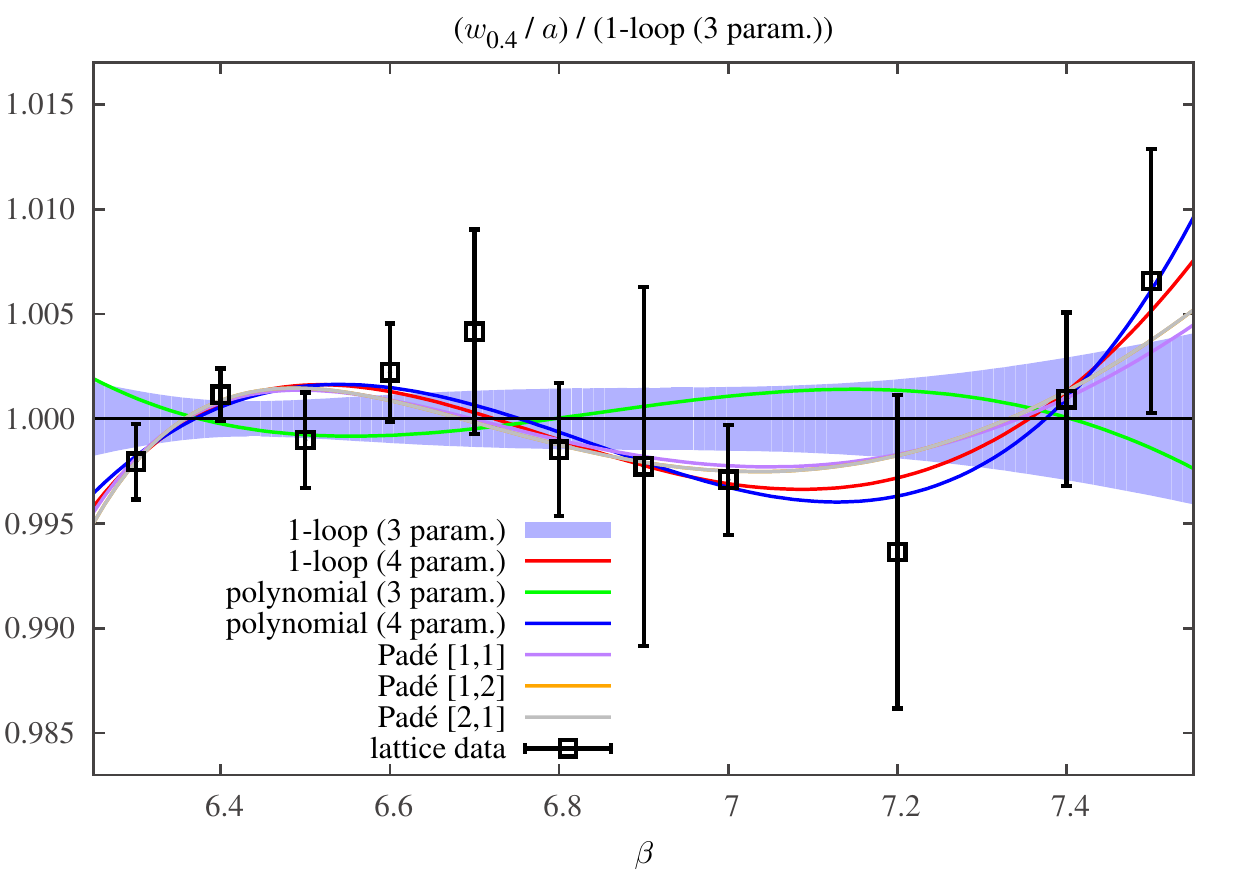}
\end{center}
\caption{
Result of the three parameter fit of~$w_{0.4}/a$ given
in Eq.~(3.3) \cite{Asakawa:2015vta}. Squares are the data obtained 
by the numerical analyses normalized by~Eq.~(3.3). 
Shaded band indicates the uncertainty originating from the errors 
of the fitted coefficients. Results with several other fitting 
functions normalized by~Eq.~(3.3) are also plotted
as well; see, Ref.~\cite{Asakawa:2015vta}.
}
\label{fig:fit}
\end{figure}

For practical applications, it is convenient to introduce a
parametrization of the ratio~$w_{0.4}/a$ in terms of~$\beta$. 
We have carried out such parametrization using various types of 
fitting functions.
Among them, the three parameter fit motivated by the one-loop
perturbation theory provides a reasonable result
($\chi^2/{\mathrm{dof}}=0.917$) for $11$ data points 
in~$6.3\le\beta\le7.5$ without over fitting:
\begin{equation}
   \frac{w_{0.4}}{a}
   =\exp\left(\frac{4\pi^2}{33}\beta
   -8.6853
   +\frac{37.422}{\beta}
   -\frac{143.84}{\beta^2}
   \right)
   [1\pm0.004(\mathrm{stat.})\pm 0.007(\mathrm{sys.})] .
\label{eq:fitfinal}
\end{equation}

In  Fig.~\ref{fig:fit}, we show the numerical results of 
$w_{0.4}/a$ normalized by the fitting function
Eq.~(\ref{eq:fitfinal}). 
The shaded band in~Fig.~\ref{fig:fit} is the error
associated with the fitting parameters in~Eq.~(\ref{eq:fitfinal}). 
The results of some other fitting functions normalized
by~Eq.~(\ref{eq:fitfinal}) are also plotted in Fig.~\ref{fig:fit}
\cite{Asakawa:2015vta}. 
They agree with each other within~$0.5\%$ in the range, 
$6.3\le\beta\le7.5$. 

We have checked that our parametrization agrees with the previously 
known ones in the range of $\beta$ at which both parametrizations
are applicable within the error \cite{Asakawa:2015vta}.
In Ref.~\cite{Asakawa:2015vta}, we have also determined 
the relation between $w_{0.4}$ and other reference scales, 
such as $t_0=t_{0.3}$, $w_0=w_{0.3}$, the Sommer scale $r_0$, 
lambda parameter and critical temperature.

\section{Thermodynamics}

Next, we report on the update of the measurement of 
thermodynamics performed in Refs.~\cite{Asakawa:2013laa,Kitazawa:2014uxa}.
In this analysis, we use the energy-momentum tensor (EMT)
defined by 
the small flow time expansion \cite{Luscher:2011bx,Suzuki:2013gza}.
This expansion asserts that 
a composite operator $\tilde{O}(t,x)$ at positive flow time
in the $t\to0$ limit can be written by a superposition of 
operators of the original gauge theory at $t=0$ as 
\begin{align}
  \tilde{O}(t,x) \xrightarrow[t\to0]{} \sum_i c_i(t) O_i^{\rm R}(x) ,
  \label{eq:SFTE}
\end{align}
where $O_i^{\rm R}(x)$ on the right-hand side represents renormalized 
operators in some regularization scheme in the original gauge 
theory at $t=0$ with the subscript $i$ denoting different operators,
and $x$ represents the coordinate in four-dimensional space-time.
Similarly to the Wilson coefficients in the operator product 
expansion (OPE), the coefficients $c_i(t)$ in Eq.~(\ref{eq:SFTE}) 
can be calculated perturbatively \cite{Luscher:2011bx}.

Using Eq.~(\ref{eq:SFTE}), one can define the renormalized 
EMT \cite{Suzuki:2013gza}.
For this purpose, we first consider the operators defined in 
Eqs.~(\ref{eq:E}) and (\ref{eq:U}),
which are dimension-four gauge-invariant operators,
as the left-hand side in Eq.~(\ref{eq:SFTE}).
Although these operators are quite similar 
to the trace and traceless-part of the EMT, they are not 
the EMT since $G_{\mu\nu}(t,x)$ is defined at nonzero 
flow time $t>0$.
Because these operators are gauge invariant,
when they are expanded as in Eq.~(\ref{eq:SFTE})
only gauge invariant operators can appear in the 
right-hand side.
Such operator with the lowest dimension is an identity operator.
In the expansion of the traceless operator Eq.~(\ref{eq:U}), 
however, the constant term cannot appear.
The next gauge-invariant operators are the dimension-four EMTs.
Up to this order, therefore, the small flow time expansions
of Eqs.~(\ref{eq:U}) and (\ref{eq:E}) are given by 
\begin{align}
   U_{\mu\nu}(t,x)
   &=\alpha_U(t)\left[
   T_{\mu\nu}^R(x)-\frac14 \delta_{\mu\nu}T_{\rho\rho}^R(x)\right]
   +O(t),
\label{eq:(2)}\\
   E(t,x)
   &=\left\langle E(t,x)\right\rangle_0
   +\alpha_E(t)T_{\rho\rho}^R(x)
   +O(t),
\label{eq:(3)}
\end{align}
where $\langle\cdot\rangle_0$ is vacuum expectation value and 
$T_{\mu\nu}^R(x)$ is the correctly normalized conserved EMT 
with its vacuum expectation value subtracted. Abbreviated are 
the contributions from the operators of dimension~$6$ or higher, 
which are proportional to powers of $t$ because of dimensional 
reasons, and thus suppressed for small $t$.

Combining relations Eqs.~(\ref{eq:(2)}) and~(\ref{eq:(3)}), we have
\begin{align}
   T_{\mu\nu}^R(x)
   =\lim_{t\to0}\left\{\frac{1}{\alpha_U(t)}U_{\mu\nu}(t,x)
   +\frac{\delta_{\mu\nu}}{4\alpha_E(t)}
   \left[E(t,x)-\left\langle E(t,x)\right\rangle_0 \right]\right\}.
\label{eq:T^R}
\end{align}
The coefficients $\alpha_U(t)$ and $\alpha_E(t)$ are calculated
perturbatively up to next to leading order in Ref.~\cite{Suzuki:2013gza}.
Using Eq.~(\ref{eq:T^R}), the energy density $e$ and pressure $p$
are obtained by taking the expectation values of diagonal components
as
\begin{align}
e= \langle T_{00} \rangle, \quad 
p=\frac13 \langle T_{11} + T_{22} + T_{33} \rangle.
\end{align}

\begin{table}[t]
\begin{center}
\begin{tabular}{|c||c|c|c|c|c|c|c|c|c|c|c|}
\hline $N_t$ & 
12    & 16    & 20    & 24    & 32 \\
\hline $\beta$ & 
6.719 & 6.941 & 7.117 & 7.256 & 7.500 \\
\hline
\end{tabular}
\caption{
The simulation parameter for the measurement of thermodynamics 
at $T=1.66T_c$.}
\label{table:param2}
\end{center}
\end{table}

\begin{figure}[t]
\begin{center}
\includegraphics[width=0.47\textwidth]{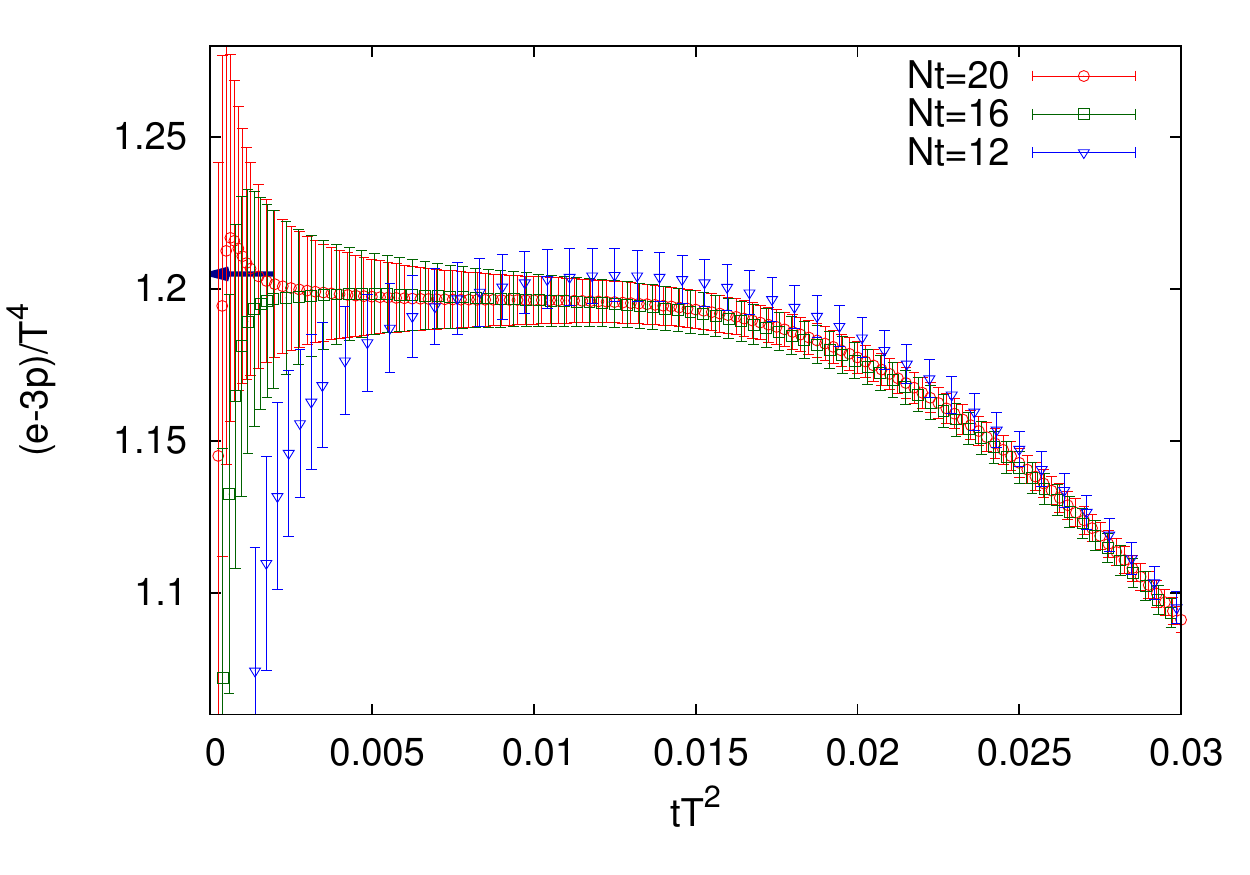}
\includegraphics[width=0.47\textwidth]{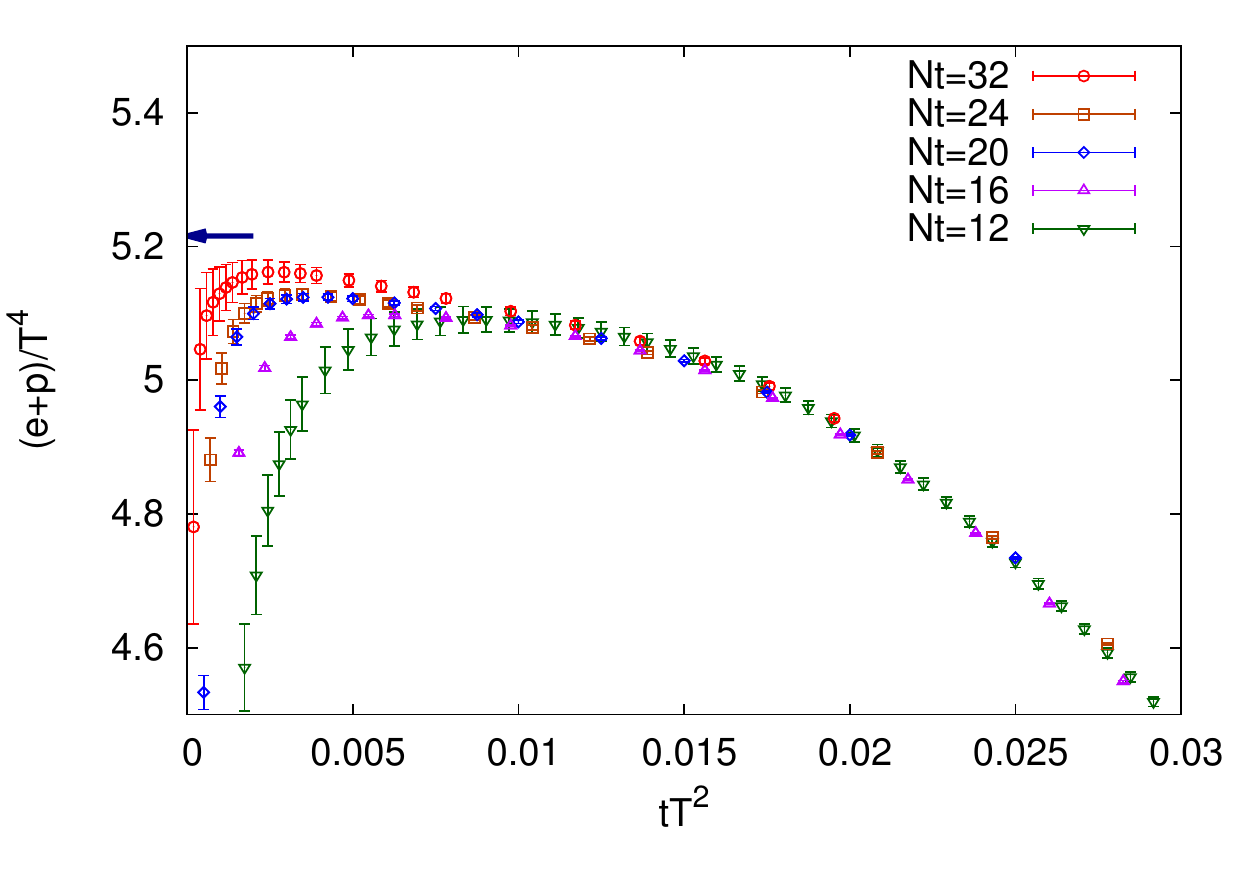}
\end{center}
\caption{
Flow time dependence of the dimensionless interaction measure 
$(e-3p)/T^4$ (left panel) and the dimensionless entropy density 
$(e+p)/T^4$ (right panel) for different
lattice spacings at $T/T_c=1.66$.
The continuum extrapolated result obtained in the integral method
in Ref.~\cite{Borsanyi:2012ve} is indicated by the arrow at vertical axis.
}
\label{fig:166}
\end{figure}

In the following, we present our numerical results 
on the thermodynamics with $N_t=12$ -- $32$ for $T=1.66T_c$
of SU(3) gauge theory with Wilson gauge action.
The values of $\beta$ for different temporal lattice length $N_t$
are determined using Eq.~(\ref{eq:fitfinal}).
These values are shown in Table~\ref{table:param2}.
All lattices have the aspect ratio $N_s/N_t$ larger than $5.3$.
We use the Wilson gauge action~$S_{\mathrm{YM}}$ for the flow equation
in~Eq.~(\ref{eq:GF}). The operators $E$ and $U$ on the lattice
are constructed from the clover-type representation of~$G_{\mu\nu}^a$.

In Fig.~\ref{fig:166}, we show the numerical results for the 
dimensionless trace anomaly $\Delta/T^4=(e-3p)/T^4$ and the 
dimensionless entropy density $s/T^3=(e+p)/T^4$ 
at~$T=1.66T_c$ as functions of the flow parameter~$tT^2$.
The figure shows that these functions have a linear behavior
in the moderate values of $tT^2$.
The numerical results show deviation from this trend 
for small and large $t$.
The deviation at small $t$ in the range $\sqrt{8t}\lesssim2a$
is attributed to the lattice discretization effects.
On the other hand, the decrease at large $t$ in the range 
$\sqrt{8t}\gtrsim1/(2T)$ is understood as the oversmearing;
the smearing by the gradient flow exceeds the temporal lattice size
in this range \cite{Asakawa:2013laa}.
In both panels in Fig.~\ref{fig:166}, we show the 
continuum-extrapolated values of $\Delta/T^4$ and $s/T^3$ with 
the same temperature obtained by the integral method \cite{Borsanyi:2012ve}
by the blue arrow beside the $y$ axis.
The figure shows that the $y$-intercepts of the linear behavior 
of our results well agree with this result.
To extract the physical values of $\Delta/T^4$ and $s/T^3$ in 
our method, we have to take the double limit $(a,t)\to(0,0)$ 
from the range satisfying $\sqrt{8t_{\rm min}}\gg2a$.
This analysis will be reported in the future publication.

The measurement of thermodynamic quantities using gradient flow
is also applicable to full QCD \cite{Makino:2014taa}. 
A first numerical analysis is reported in Ref.~\cite{Itou}.

Numerical simulation for this study was carried out on 
IBM System Blue Gene Solution at KEK under its Large-Scale 
Simulation Program (Nos.~T12-04, 13/14-20 and 14/15-08). 
The work of M.~A., M.~K., and H.~S. are supported in part by a 
Grant-in-Aid for Scientific Researches~23540307 and 26400272, 
25800148 and 23540330,respectively. 
E.~I. is supported
in part by Strategic Programs for Innovative Research (SPIRE) Field~5. 
T.~H. is partially supported by RIKEN iTHES Project.

\end{document}